\def\year{2022}\relax
\documentclass[letterpaper]{article} % DO NOT CHANGE THIS
\usepackage{aaai22}  % DO NOT CHANGE THIS
\usepackage{times}  % DO NOT CHANGE THIS
\usepackage{helvet}  % DO NOT CHANGE THIS
\usepackage{courier}  % DO NOT CHANGE THIS
\usepackage[hyphens]{url}  % DO NOT CHANGE THIS
\usepackage{graphicx} % DO NOT CHANGE THIS
\usepackage{natbib}  % DO NOT CHANGE THIS AND DO NOT ADD ANY OPTIONS TO IT
\usepackage{caption} % DO NOT CHANGE THIS AND DO NOT ADD ANY OPTIONS TO IT
\DeclareCaptionStyle{ruled}{labelfont=normalfont,labelsep=colon,strut=off} % DO NOT CHANGE THIS
\usepackage{algorithm}
\usepackage{algorithmic}

\usepackage{comment}
\usepackage{tabularx}
\usepackage[normalem]{ulem} % added in the new revision
\usepackage{xcolor, soul} % added in the new revision
\sethlcolor{green}
\usepackage{soul}
\usepackage{graphicx}
\usepackage{mathrsfs}
\usepackage{amsmath}
\usepackage{placeins}
\usepackage{epstopdf, epsfig}
\usepackage{subcaption}
\usepackage[export]{adjustbox}
\usepackage{booktabs}
\usepackage{amsfonts}

\usepackage{dcolumn}% Align table columns on decimal point
\usepackage{bm}% bold math
\newcommand{\upi}{\pi}

\usepackage{newfloat}
\usepackage{listings}
\floatstyle{ruled}
\newfloat{listing}{tb}{lst}{}
\floatname{listing}{Listing}
\title{Emulating Spatio-Temporal Realizations of Three-Dimensional Isotropic Turbulence via Deep Sequence Learning Models}
\author{
    Mohammadreza Momenifar\textsuperscript{\rm 1},
    Enmao Diao\textsuperscript{\rm 2},
    Vahid Tarokh\textsuperscript{\rm 2},
    Andrew D. Bragg\textsuperscript{\rm 1},
}
\affiliations{
    \textsuperscript{\rm 1}Department of Civil and Environmental Engineering, 
    \textsuperscript{\rm 2}Department of Electrical and Computer Engineering\\
    Duke University\\
    Durham, NC, 27701, USA\\
    mohammadreza.momenifar@duke.edu, 
    enmao.diao@duke.edu, 
    vahid.tarokh@duke.edu, 
    andrew.bragg@duke.edu
}

\usepackage{bibentry}
\begin{document}

\maketitle

\begin{abstract}
	We use a data-driven approach to model a three-dimensional turbulent flow using cutting-edge Deep Learning techniques. The deep learning framework incorporates physical constraints on the flow, such as preserving incompressibility and global statistical invariants of velocity gradient tensor. The accuracy of the model is assessed using statistical and physics-based metrics. The data set comes from Direct Numerical Simulation of an incompressible, statistically stationary, isotropic turbulent flow in a cubic box. Since the size of the dataset is memory intensive, we first generate a low-dimensional representation of the velocity data, and then pass it to a sequence prediction network that learns the spatial and temporal correlations of the underlying data. The dimensionality reduction is performed via extraction using Vector-Quantized Autoencoder (VQ-AE), which learns the discrete latent variables. For the sequence forecasting, the idea of Transformer architecture from natural language processing is used, and its performance compared against more standard Recurrent Networks (such as Convolutional LSTM). These architectures are designed and trained to perform a sequence to sequence multi-class classification task in which they take an input sequence with a fixed length (k) and predict a sequence with a fixed length (p), representing the future time instants of the flow.
	Our results for the short-term predictions show that the accuracy of results for both models  deteriorates  across  predicted  snapshots  due  to  autoregressive  nature  of  the predictions. Based  on  our  diagnostics tests, the trained Conv-Transformer model outperforms the Conv-LSTM one and can accurately, both quantitatively and qualitatively, retain the large scales and capture well the inertial scales of flow but fails at recovering the small and intermittent fluid motions. 

\end{abstract}
\vspace{-0.2cm}
\section{Introduction}
\vspace{-0.1cm}
Turbulence is a complex dynamical system which is strongly high-dimensional, multi-scale, non-linear, non-local and chaotic with a broad range of correlated scales that vary over space and time. Such features make high-fidelity spatio-temporal simulation of turbulence extremely challenging and often impossible (particularly for large domains, unsteady flows, complex boundary conditions) due to limitations of computational power and numerical schemes.
From Computational Fluid Dynamics (CFD) perspective, turbulent flows can be simulated by solving Navier-Stokes (NS) equations in three flavors of Direct Numerical Simulation (DNS), Large Eddy Simulation (LES) and Reynolds Averaged Navier-Stokes Simulation (RANS) depending on the application and expected accuracy. 
In DNS, full-scale resolution is achieved by solving NS on a domain with extremely fine spatial and temporal grids so that all the scales down to the dissipation range are resolved.
The LES is based on low-pass spatial filtering of NS equations in which the small scales of turbulent flow are modeled and large unsteady motions, corresponding to the most energetic scales, are resolved. The main goal of modeling in LES is to build a more accurate and universal closure model for the residual (sub-grid) stress tensor.
RANS models are derived by taking a temporal average of NS equations, resulting in simpler steady equations, and assuming a linear relationship between Reynolds stresses and mean strain rate. In RANS modeling, the focus is on constructing the Reynolds-stress tensor closure. 

The aforementioned CFD techniques are purely physics based where NS equations are solved using different differentiation and integration schemes. However, the high-fidelity simulations using DNS and LES are computationally prohibitive and are limited to simplified turbulent flows to gain detailed insight into the physics of turbulence. Furthermore, RANS cannot simulate important characteristics of turbulence such as unsteadiness and intermittency due to its time averaging nature. Therefore, a major pursuit and fundamental task of turbulence community is to develop new modeling techniques to characterize dynamics of turbulence that evolve over time and space while varying over a broad range of spatio-temporal scales.  

Recently with growing the availability of high-fidelity data and computational power data-driven modeling has gained huge interest and been introduced as a competitive alternative to conventional numerical simulations. Among these data-driven methods, Machine Learning (ML) techniques, particularly Deep Learning (DL) models, have received more attention due to their ability to capture complex interactions and achieve outstanding performance across a wide range of applications in information technology, healthcare and engineering to name a few.
%Deep learning models, also known as deep neural networks, are built from stack of layers with learnable parameters that perform linear and non-linear transformation between their inputs and outputs where their parameters are trained through an optimization procedure.

%Despite remarkable success of Deep Neural Networks and their outperformance against competitive algorithms in many areas of big data analytics tasks, their applications in physical sciences have been tempered by the caveat that these networks are treated as "black boxes" which are hard to interpret and physics agnostic. Furthermore, training such networks requires a lot of data (except for unsupervised learning tasks), which may not be available, and the training process takes a fair amount of time. However, well-trained deep learning models can make quick inferences about similar input data, particularly for very complex physical systems, and this characteristic distinguishes them from conventional numerical solvers.

In this research our main objective is to bring fresh perspective to the classic problem of turbulence modeling by exploring and utilizing modern deep learning techniques to develop accurate and efficient turbulence models. More specifically, the primary goal is to develop a data-driven modeling framework to simulate high-fidelity three-dimensional turbulent flow realizations that respect turbulence characteristics without solving the full NS equations.

%This outline of this paper is given next. A  brief literature review on the data compression techniques, application of deep learning in turbulence research and data compression is provided in Section \ref{Literature}. In Sections \ref{Methodology} our mathematical methodology, and in  Section \ref{Computational_Details} our implementation approach is discussed.  Assessment metrics and methodologies are described in Section \ref{Evaluation_Metrics}. Using these metrics, numerical results are provided  in Section \ref{Results_Discussion} evaluating the performance of our model on several test cases. These test cases represent turbulent flows with different characteristics from the training set data. Finally, Section \ref{Conclusions} provides a brief summary of our findings.
\vspace{-0.2cm}
\section{Background}
\vspace{-0.1cm}
\label{Literature}
In fluid dynamics community, particularly computational fluid dynamics of turbulent flows, there have been some attempts in recent years to utilize deep learning tools into turbulence modeling and generate predictive models. These endeavors have been mainly centered around developing Reynolds stress closures and subgrid-scale (SGS) models for RANS and LES simulations respectively \cite{maulik2019subgrid,ling2016reynolds,beck2019deep}, super-resolution reconstruction (enhancing the resolution) of coarse flow fields \cite{fukami2018super}, turbulence data compression \cite{glaws2020deep}, and augmenting existing turbulence models with physics-informed machine learning \cite{wu2018physics,wang2017origin}.

The first step of modeling dynamics of turbulence is to identify potential deep learning models which are well-suited for handling non-linear spatial and sequential data. Indeed, we seek models that can capture spatial and temporal dependencies of turbulent flow field.
Given the grid/pixel-based discretizations of our computational domain, we can draw an analogy between turbulence data and image data where three components of velocity field represent RGB (red, green, blue) color channels and instead of two dimensional spatial grids we have three-dimensional ones. Therefore, we can benefit from modern computer vision approaches in our modeling framework, particularly Convolutional Neural Networks (CNN) \cite{khan2018guide}.

The capability of CNNs in extracting spatial distribution of data (exploiting the correlations between the adjacent input data points) could make them a suitable architecture for a variety of physics-based applications. %Furthermore, one can incorporate different boundary conditions (wall, periodic etc.) into convolutional layers in the form of padding \cite{patil2019development}. 
In the past few years there have been multiple studies that utilized CNNs in the context of spatio-temporal modeling of turbulence \cite{wang2020towards,li2020fourier} and found promising results. However, these works have been limited to 2D turbulent flows which are far less complicated than 3D turbulence both in terms of physics and computational tractability. The only relevant works aligned with our objective are the recent studies of \cite{mohan2019compressed, mohan2020spatio, mohan2020wavelet} that their findings, though encouraging but seem suspicious.

In \cite{mohan2020spatio}, the dynamic mapping of ScalarHIT dataset, containing three components of velocity field of isotropic turbulence and two passive scalar advected  with the flow, was modelled via a integration of Convolutional Autoencoder neural network (CAE) and convolutional LSTM, so called Compressed Convolutional LSTM (CCLSTM). In this architecture, flow snapshots are first transformed to a low-dimensional latent-space, using a pre-trained CAE, serving as input sequence of the sequence learning model. This LSTM model is trained to predict an output sequence representing the future snapshots of flow in the latent space and then this output sequence is transformed to original dimension of flow field. This CC-LSTM model is autoregressive (cycling prediction), meaning that previous predictions are fed as the input for the next predictions to generate future temporal realizations. Their CAE has a compression ratio of 125 %(corresponding to 125-fold decrease in the size of flow snapshot). 
and the temporal spacing between sampled snapshots, or sampling rate, was $\omega = 0.09 \: \tau$, where $\tau$ represents eddy turnover time.

The static reconstruction results show that their CAE model has retained large %scales and has some discrepancies in 
and inertial scales but fails drastically at capturing small scales, as demonstrated in the tails of the PDF of velocity gradient and large wave numbers in energy spectra (Figure 10 in \cite{mohan2020spatio}). For the dynamic mapping, they predicted several time instants over range $\tau = [3. 4.5]$. Their (dynamic) results (Figure 16 in \cite{mohan2020spatio}) show that their model performs even better than the static reconstructions and can fully capture large and inertial scales of flow and generate stable flow realizations over this $1.5 \: \tau$ prediction horizon. These results are really suspicious as we observe that the tails of PDF of velocity gradient, which their compression model failed to retain, are almost perfectly recovered and remained stable during dynamic prediction. The authors conjectured that their model adds some artifacts to the predicted snapshots that result in mimicking these intermittent regions of the PDFs. More importantly, their claim (and results) that the error of their temporal predictions remains marginal with increasing prediction horizon is in clear contrast with the findings reported in the recent spatio-temporal modeling of two-dimensional turbulence\cite{wang2020towards}.

%the symmetric nature of PDFs indicates that the network may be generating some random noise to compensate for information loss in the small scales

The same authors in a recent work \cite{mohan2020wavelet}, proposed another spatio-temporal model, called Wavelet-CLSTM, in which the compression is performed via wavelet transformation as opposed to CAE. Their results again indicate that such a model can emulate temporal evolution of flow field such that statistics of large and intermediate scales are retained stable with increasing prediction horizon. While it was clear that their model has large discrepancies in recovering the PDF of velocity gradient, they reported that these PDFs solely test the smallest scales of flow and so such deviations are expected since the loss of information (during wavelet transform) happens at small scales. However, from turbulence literature we know that PDFs of velocity gradient tensor also provide rich multi-scale information and they do not only represent small scales of flow.

% \subsection{Data Compression}

% \subsection{Deep Learning in Turbulence Research}

% \subsection{Deep Learning for Data Compression}
\vspace{-0.2cm}
\section{Method}\label{Method}
\vspace{-0.1cm}
In this research, we leverage existing high-fidelity DNS data to emulate spatio-temporal evolution of three-dimensional turbulence with a less memory and computation costs compared to existing flow solvers. More importantly, we make this framework physics-informed through embedding a calibration inside model by infusing as much as our prior knowledge of data and turbulence as possible into training process. %Our hope is to advocate this deep learning data-driven physics-informed modeling framework, upon its success, as a promising compensate for traditional turbulence modeling techniques. 
Hierarchical superposition of complex structures in deep convolutional layers resemble cascade nature of turbulence and hence it is worth exploring whether the techniques utilized in computer vision tasks could be applicable to turbulent flow field. Therefore, the convolutional layers would be the core part of the deep learning architecture of this study. We probe and validate this framework by employing high-fidelity DNS data from three-dimensional statistically stationary isotropic turbulence.  

The high-fidelity DNS data come from solving the NS equations on a fine-grid three-dimensional mesh. In this study, we start with our smallest dataset which represents a fully resolved statistically stationary isotropic turbulent flow with $R_\lambda = 90$ simulated on a cubic domain with 128 grid point in each direction. At each of these grid points, we have three components of velocity field ($u , v, w$) and nine components of velocity gradient tensor defined as $A_{ij}=\nabla \boldsymbol{u}=\partial u_i/\partial x_j$, in which $u_i$ is a component of the fluid velocity, and $x_i$ is a spatial coordinate. 

Since the size of the dataset is memory intensive, similar to the conceptual design of \cite{mohan2019compressed}, we first generate a low-dimensional representation of the velocity data and then pass it to a sequence prediction network that learns the positional and temporal correlations of the underlying data. 
Therefore, our framework will be composed of two separate models where one serves as a compression engine and the other performs prediction. In the deployment phase, once both engines are trained and we want to test the performance of our framework, a sequence of input data is first compressed via the down-sampling part of the compression network, then the prediction network takes this collection of reduced representations and outputs a sequence of future time steps in the latent space and finally this sequence is mapped to original size representation via the upsampling part of the compression engine. %It is worth mentioning that achieving the appropriate/optimal architecture of the framework demands thorough ablation study and several trials of hyper-parameter (over their search domain).

\textbf{Vector-Quantized Autoencoder (VQ-AE)}\; The dimensionality reduction is performed via extraction using a variant of autoencoder (AE) network, called Vector-Quantized Autoencoder (VQ-AE) architecture \cite{momenifar2021dimension, van2017neural, razavi2019generating}, which encodes the input data in a discrete latent space and can effectively use the capacity of latent space by conserving important features of data that usually span many dimensions in data space (such as objects in images) and reducing entropy (putting less focus on noise) \cite{van2017neural}. Mathematically speaking, in a vector quantization (VQ) operation, $m$-dimensional vectors in ${\mathbb{R}^m}$ are mapped into a finite set of codewords/vectors
$\mathcal{Y}=\{e_{i}:i=1,2,..,K\}$ with a fixed size $D$ or $e_{i}\in {\mathbb{R}^D}$, where $K$ represents the size of the codebook.
Compared to a conventional autoencoder, a Vector-Quantized Autoencoder has an additional Vector-Quantizer module. The encoder ($E$) serves as a non-linear function that maps input data ($x$) to a vector $E(x)$. The quantizer modules takes this vector and outputs an index ($k$) corresponding to the closest codeword in the codebook to this vector ($e_{k}$):
\begin{align}\label{eq:Quantizer}
\text{Quantize}(E(x)) = e_{k},\, k = \underset{j}{\arg\min} \parallel E(x) - e_{j} \parallel_{2}.
\end{align}

Codeword index $k$ is used for the integer representation of the latent space, and $e_{k}$ serves as the input of decoder ($D$) which operates as another non-linear function to reconstruct the input data. The Vector-Quantizer module brings two additional terms in the loss function, namely codebook loss and commitment loss, to align the encoder output with the vector space of the codebook. The entire VQ-AE loss is defined as:
\begin{align}\label{eq:VQ_VAE_Loss}
\mathcal{L}(x,D(e))=\underbrace{\parallel x - D(e) \parallel^{2}_{2}}_{reconstruction~loss} + \nonumber \\
\underbrace{\parallel sg\{E(x)\} - e \parallel^{2}_{2}}_{codebook~loss} + \underbrace{\beta \parallel sg\{e\} - E(x) \parallel^{2}_{2}}_{commitment~loss}. 
\end{align}
As noted earlier, preserving small-scale properties of the turbulent flow was a challenge for prior compression models. Here we add appropriate constraints in order to capture these more faithfully. More details on the properties of these physics-based constraints can be found in \cite{momenifar2021dimension}. By adding these constraints as regularization terms to the VQ-AE loss function gives the overall loss function (OL) given below
\begin{align}\label{eq:Final_loss}
\text{Overall Loss (OL) } = \text{VQ-AE loss} \nonumber \\ 
+ \alpha \times \text{VGC} + \gamma \times \text{OC}
\end{align}
\begin{align}\label{eq:VG_Constraint}
\text{Velocity Gradient Constraint (VGC)} = \underbrace{MSE(A_{ij},\widehat{A_{ij}})}_{i= j} + \nonumber \\ 
a \times \underbrace{MSE(A_{ij},\widehat{A_{ij}})}_{i\neq j} \nonumber \\
\end{align}
\begin{align}\label{eq:Other_constraints}
%\begin{split}
\text{Other Constraints (OC)} =  MAE(\langle S_{ij} S_{ij} \rangle,\widehat{\langle S_{ij} S_{ij} \rangle}) +\nonumber \\ 
MAE(\langle R_{ij} R_{ij} \rangle,\widehat{\langle R_{ij} R_{ij} \rangle}) + \nonumber \\
MAE(\langle S_{ik} S_{kj} S_{ij} \rangle,\widehat{\langle S_{ik} S_{kj} S_{ij} \rangle})+ \nonumber \\
MAE(\langle S_{ij} \omega_{i} \omega_{j} \rangle,\widehat{\langle S_{ij} \omega_{i} \omega_{j} \rangle}), 
%\end{split}
\end{align}
% \subsection{Convolutional LSTM}
\textbf{Convolutional LSTM}\; Vanilla RNN is known to suffer from gradient vanishing and explosion problems, and it fails to capture long term dependencies among sequential data. Therefore, Long-Short Term Memory (LSTM) is later proposed by \cite{hochreiter1997long} as a solution to these drawbacks by introducing memory state and multiple gating mechanism. As formulated below, LSTM has an internal memory cell $c_t$ to store the long term information. It also requires four gates to control the information flow from input, hidden state and the memory cell. The input gate $i_t$ determines how much information of input and hidden state should be remembered by the memory cell. The forget gate $f_t$ determines how much long term memory should be saved for the next time step. $g_t$ contains the information of current input and previous hidden state and the output gate $o_t$ controls the information flows into the next hidden state. Depending on the switching of gates, LSTM can represent long-term and short-term dependencies of sequential data simultaneously.
\begin{align}
i_t &= \sigma(W_{xi}x_t + b_{xi} + W_{hi}h_{t-1} + b_{hi}) \\
f_t &= \sigma(W_{xf}x_t + b_{xf} + W_{hf}h_{t-1} + b_{hf}) \\
g_t &= \tanh(W_{xg}x_t + b_{xg} + W_{hg}h_{t-1} + b_{hg})  \\ 
o_t &= \sigma(W_{xo}x_t + b_{xo} + W_{ho}h_{t-1} + b_{ho}) \\
c_t &= f_t*c_{t-1}+i_t*g_t \\
h_t &= o_t*\tanh(c_t)
\end{align}
Apart from LSTM mentioned above, there still exist many other variations of RNN. Many of them are designed to specialize on specific type of sequential data \cite{cho2014learning, jozefowicz2015empirical,tai2015improved,greff2016lstm,kent2019performance, diao2019restricted}. Because we focus on three-dimensional turbulence data, we use Convolutional LSTM (Conv-LSTM)\cite{xingjian2015convolutional} to model the dependence between the spatial and temporal information. Conv-LSTM replace linear transformation in vanilla LSTM with Convolutional Neural Networks (CNN) and its general formulation remains the same.

% \subsection{Convolutional Transformer}
\textbf{Convolutional Transformer}\; Transformer \cite{vaswani2017attention} architecture has been introduced with the purpose of offering parallel computation (by avoiding recursion that consequently reduces training time) and reducing performance drop due to long-term dependency issues. The transformer model introduces the self-attention unit, which accounts for similarity scores between elements of a sequence, and positional embeddings, a unit that replaces the recurrence. These innovative units can capture sequential relationship between different items of a sequence and consequently allow the Transformer network to process input sequences as a whole rather than element by element, which is typical in recurrent neural networks. This characteristic, processing all the items in an input sequence simultaneously, enables the Transformer model not to rely on previous hidden states for preserving correlations with previous elements in sequence (no backpropagation through time), hence eliminates the risk of forgetting past information with increasing sequence length. 

We formulate the transformer block below. Input $x$ is transformed into query, key, and value with three separate weight matrix $W_{q}$, $W_{k}$, and $W_{v}$ respectively. We calculate self-attention $\alpha$ by taking the dot product of the query with the key with a scaling factor $\frac{1}{\sqrt{d_k}}$ where $d_k$ is the size of hidden representations. The self-attention is used to attend the value with a linear transformation parameterized by weight matrix $W_u$. We use a feed-forward block parameterized by weight matrix $W_1$ and $W_2$ to further model the attended output. We standardize temporal information with LayerNorm \cite{ba2016layer} function. The residual connection is also used to avoid the issue of gradient vanishing.
\begin{align}
Q\left(x\right)&=W_{q}*x, K\left(x\right)=W_{k}*x, V\left(x\right)=W_{v}*x\\
\alpha&=\operatorname{softmax}\left(\frac{\left\langle Q\left(x\right), K\left(x\right)\right\rangle}{\sqrt{d_k}}\right)\\
u^{\prime}&= W_{u}*\alpha V\left(x\right)\\
u&=\text { LayerNorm }\left(x+u^{\prime}\right)\\
z^{\prime}&=W_{2}*\operatorname{ReLU}\left(W_{1}*u\right)\\
z&=\operatorname{LayerNorm}\left(u+z^{\prime}\right)
\end{align}
This block can be stacked multiple times for better performance. Similar to Conv-LSTM, we replace linear transformation in vanilla Transformer with convolution $*$ and its general formulation remains the same.

\begin{figure*}[hbtp]
\centering
 \includegraphics[width=1\linewidth]{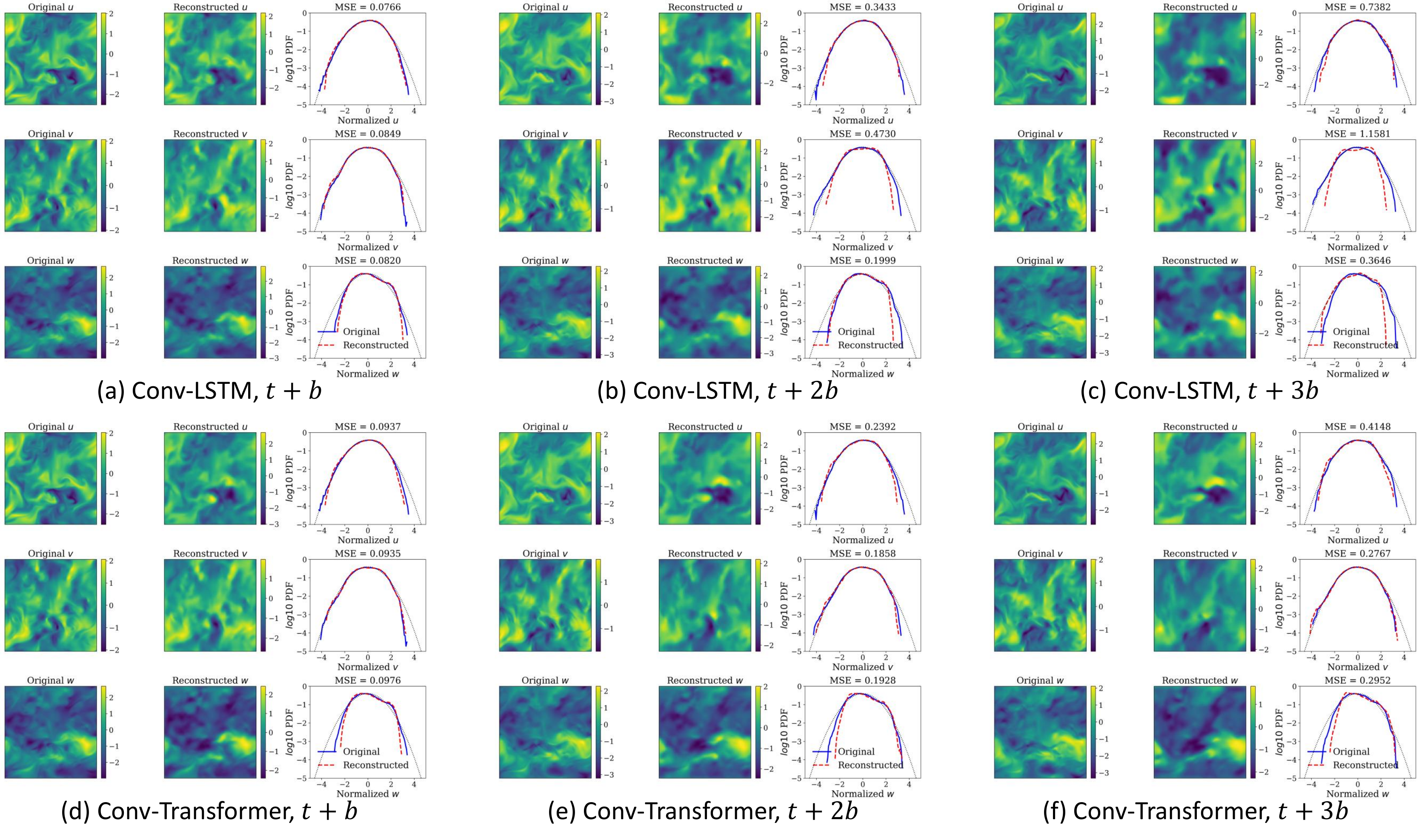}
 \caption{Predictions of velocity field using (a-c) Conv-LSTM and (d-f) Conv-Transformer models.}
 \label{fig:VelocityComponents}
\end{figure*}

\begin{figure}[hbtp]
\centering
 \includegraphics[width=1\linewidth]{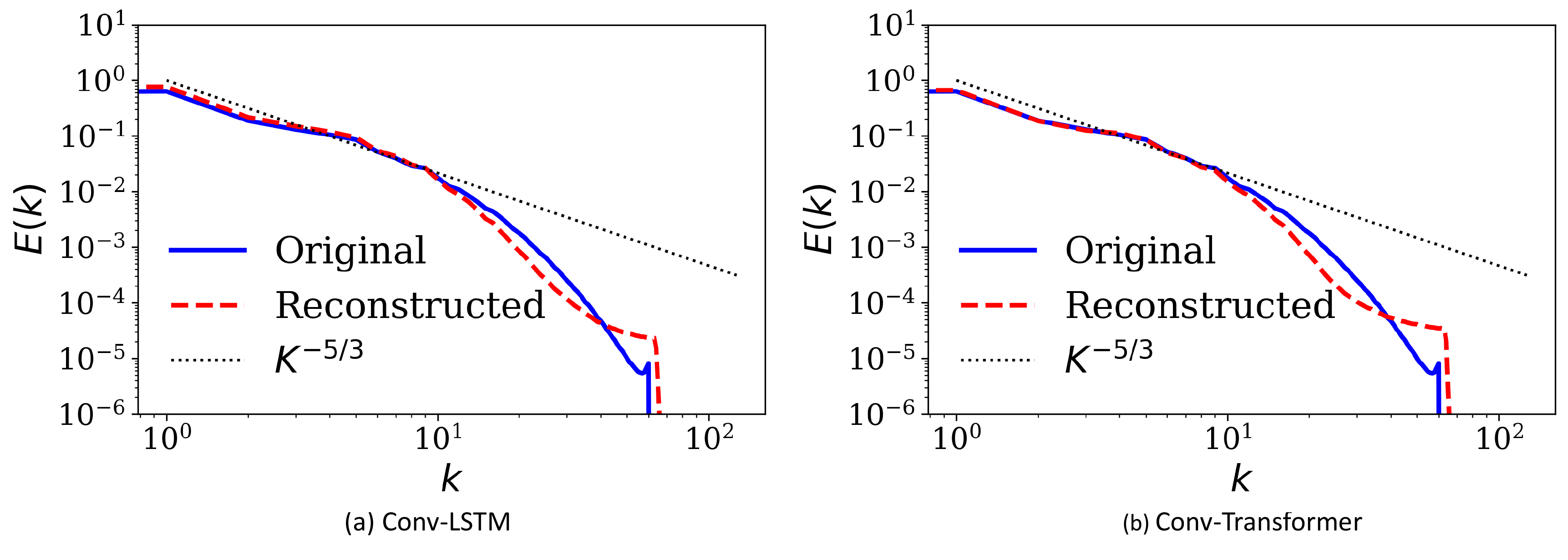}
 \caption{Predictions of turbulent kinetic energy (TKE) using (a-c) Conv-LSTM and (d-f) Conv-Transformer models.}
 \label{fig:TKE}
\end{figure}

\vspace{-0.2cm}
\section{Computational Details}\label{Computational_Details}
\vspace{-0.1cm}
As mentioned earlier, our framework consists of two deep learning models, one for compression and the other for sequence learning, which are trained separately. Our compression model is a VQ-AE, proposed in our recent work ( \cite{momenifar2021dimension}). We design our VQ-AE network so that it can downsample original data by a scaling factor of $SF = 2$. With $K = 512$ representing the size of the codebook and mapping three velocity components into one in the discrete latent space, we can achieve $\frac{3 \times 32}{1 \times 9} \times (SF)^{3}$ reduction in bits, corresponding to $85$. Indeed, an input data of shape $(3,128,128,128)$ is compressed to $(1,64,64,64)$. One can find more details regarding hyper-parameters, model architecture and training in \cite{momenifar2021dimension}. 

For the sequence learning model, we designed and trained two radically different sequence learning models, convolutional LSTM (Conv-LSTM) and convolutional Transformer (Conv-Transformer).
% It should be emphasized that there are no standard or pre-trained architectures for these models and a machine learning developer is supposed to design all the network from scratch and modify them depending on the specific applications. 
These architectures are designed and trained to take an input sequence with length $k$ and predict a sequence with length $p$, representing the next $p$ realizations of the system (sequence to sequence multi-class classification). Mathematically, it means mapping $[X_{t-k},...,X_{t-2b}, X_{t-b}, X_{t}]$ to $[X_{t+b}, X_{t+2b},..., X_{t+p}]$, where $b$ is the sampling interval (time span between observations),  $X \in {\mathbb{R}^{H \times W \times D \times C}}$ in which $C$ is the number of channels in the data (corresponding with the number of velocity components), $H$ , $W$ and $D$ represent the height, width and depth of data.
Here the architectures used in this study are briefly summarized. 

For the both sequence learning models, the input sequence of velocity field is first transformed to a low-dimensional discrete latent space, via the pre-trained encoder of the VQ-AE model, and then passes through an embedding layer to transform data to continuous space (indeed, each integer is represented with a codeword which is a vector in real space). For the Convolutional LSTM model, this continuous latent space sequence is fed to a LSTM block consisting of three LSTM cells/layers. Afterwards the output of LSTM block passes through a series of linear layers to expand the channel dimension from codeword size to the codebook one. Then we need to find, for each grid point in three-dimensional domain, the class (index) with the highest probability in this multi-class classification task. This step is performed by returning the index of the largest element in the channel dimension and then we apply the cross-entropy loss to compute the discrepancy between the true and predicted output sequence. Finally, the predicted output sequence is transformed to the original dimension of velocity field through the pre-trained decoder of the VQ-AE model. The same pipeline was used for the Transformer model, where we designed its architecture following the instructions in \cite{vaswani2017attention} and adjusted the network for image data (the Transformer model was originally proposed for text data).

In our dataset we have 80 snapshots equally spaced in time, from $ t = 3 \:T_L - 7 \:T_L$, where $T_l$ denotes large eddy turn over time. Indeed, these snapshots cover a time span of $4T_l$ corresponding to a sampling interval of $b = 0.05 \: T_L$. We use the first 40 snapshots as training set and the rest as test set.  We trained our framework for 300 epochs with batch size $=1$ using the Adam optimizer \cite{kingma2014adam} with learning rate = $0.001$, along with a learning rate scheduler. 
It should be noted that our focus has been on the general characteristics of the framework rather than achieving the best configurations, hence there might be room for improvements by tuning the proposed hyper-parameters.

\begin{figure*}[hbtp]
\centering
 \includegraphics[width=1\linewidth]{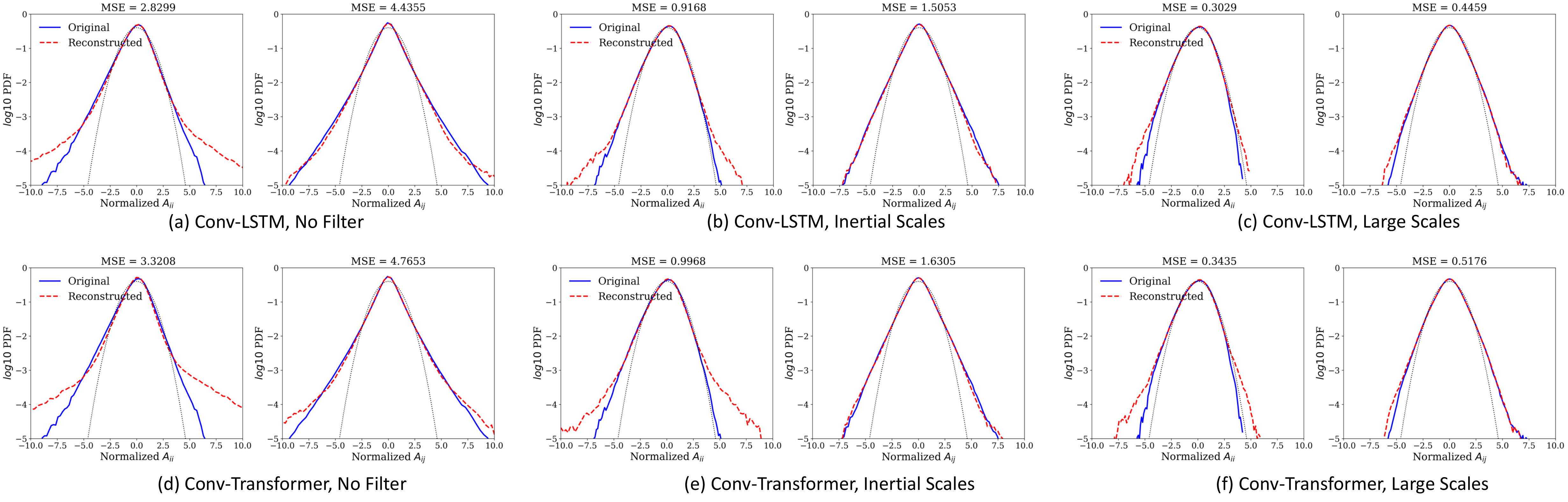}
 \caption{Predictions PDFs of velocity gradient tensor using (a-c) Conv-LSTM and (d-f) Conv-Transformer models on small (a,d), inertial (b,e), and large (d,f) scales.}
 \label{fig:VelocityGradient}
\end{figure*}

We implemented this framework in PyTorch using the CUDA environment, and trained it on one NVIDIA Pascal P100 GPU. %The training was completed within approximately 8 hours, and the maximum GPU memory consumed was around 5 $GB$.
The performance of the predicted velocity field is assessed not only via the conventional error measurement methods such as MSE and mean absolute error (MAE) but also using rigorous physics-based metrics relevant to the analysis of turbulence such as the probability density functions
(PDFs) of the filtered velocity gradient tensor and its invariants, the turbulent kinetic energy spectra and the joint PDF of $Q-R$ plane (detailed in \cite{momenifar2021dimension}).

\begin{table}[hbtp]
\centering
\label{tab:parameters}
\caption{Simulation parameters for the DNS study of isotropic turbulence (arbitrary units).
		$N$ is the number of grid points in each direction, 
		$Re_\lambda \equiv u'\lambda/\nu$ is the Taylor micro-scale
		Reynolds number, $\lambda $ is the Taylor micro-scale,
		$\mathscr{L} $ is the box size, $\nu$ is the fluid kinematic viscosity, $\epsilon$ is the mean
		turbulent kinetic energy dissipation rate,
		$l$  is the integral length scale, $\eta \equiv \nu^{3/4}/\epsilon^{1/4}$ is the Kolmogorov length scale, 
		$u' \equiv \sqrt{(2k/3)}$ is the fluid r.m.s. fluctuating 
		velocity, $k$ is the turbulent kinetic energy, 
		$u_\eta$ is the Kolmogorov velocity scale, 
		$T_L \equiv l/u^\prime$ is the large-eddy turnover
		time, $\tau_\eta \equiv \sqrt{(\nu/\epsilon)}$ is the Kolmogorov time scale, 
		$\kappa_{\rm max}=\sqrt{2}N/3$ is the maximum resolved wavenumber.}
\resizebox{1\columnwidth}{!}{
\begin{tabular}{@{}ccccccccccccc@{}}
\toprule
Parameter & $N$ & $Re_\lambda$ & $\mathscr{L} $ & $\nu$ & $\epsilon$ & $l$  & $l/\eta$ & $u'$  & $u'/u_\eta$ & $T_L$ & $T_L/\tau_\eta$ & $\kappa_{{\rm max}}\eta$ \\ \midrule
Value     & 128 & 93           & 2$\upi$        & 0.005 & 0.324      & 1.48 & 59.6     & 0.984 & 4.91        & 1.51  & 12.14           & 1.5                      \\ \bottomrule
\end{tabular}
}
\end{table}

\vspace{-0.2cm}
\section{Experiments}\label{Results_Discussion}
\vspace{-0.1cm}

% As reported in \cite{momenifar2021dimension} and shown in Table \ref{tab:Summary_HIT_SF2_Table}, 
Our VQ-AE model offers a compression ratio of $85$ and recovers all the flow characteristics up to second order statistics of the velocity gradient tensor, with small discrepancies at the smallest scales. We also found that the embedding physics constraints in the loss function can noticeably improve the quality of the reconstructed small scales of the flow. Interested readers can found details on our data compression engine in \cite{momenifar2021dimension}.

As mentioned earlier, our sequence learning models have been trained to take an input sequence with a fixed length ($k$) and predict a sequence with a fixed length ($p$), representing the future time instants of the flow. The spacing between the flow snapshots or sampling interval ($b$), is arbitrary but is constrained with the input sequence. Therefore, feeding an input sequence where all the flow realizations are $0.05T_L$ apart would result in an output sequence with a similar sampling interval $b = 0.05T_L$. In our experiments, we trained our models with sequences where $\omega = 0.05T_L$ and therefore they are well-suited to predict temporal realizations that are $0.05T_L$ apart. Furthermore, we can also roll out autoregressively (also known as cyclic prediction) and feed predicted sequence as input sequence to the model and generate flow realizations over a larger prediction horizon.

Given the size of the compressed data in discrete latent space, $(1,64,64,64)$, and our available GPU memory, we were able to train sequence learning models with $k=p=3b$. It is worth mentioning that one can train models with much larger $k, p$ with a more compressed data , corresponding to a larger $SF$. However, this would come at the cost of losing prediction accuracy as the information content of compressed data with $SF=4$ is less richer that $SF=2$.
In what follows, we present our results during inference for short-term prediction, where the sampling interval in the test data is the same as the training data ($b = 0.05T_L$).% (ii) varying sampling interval prediction, where the sampling interval in the test data is a few times larger than the training data, and (iii) long-term prediction, where multiple time instants are predicted autoregressively to assess the error propagation over a larger time span in future.

\begin{figure*}[hbtp]
\centering
 \includegraphics[width=0.8\linewidth]{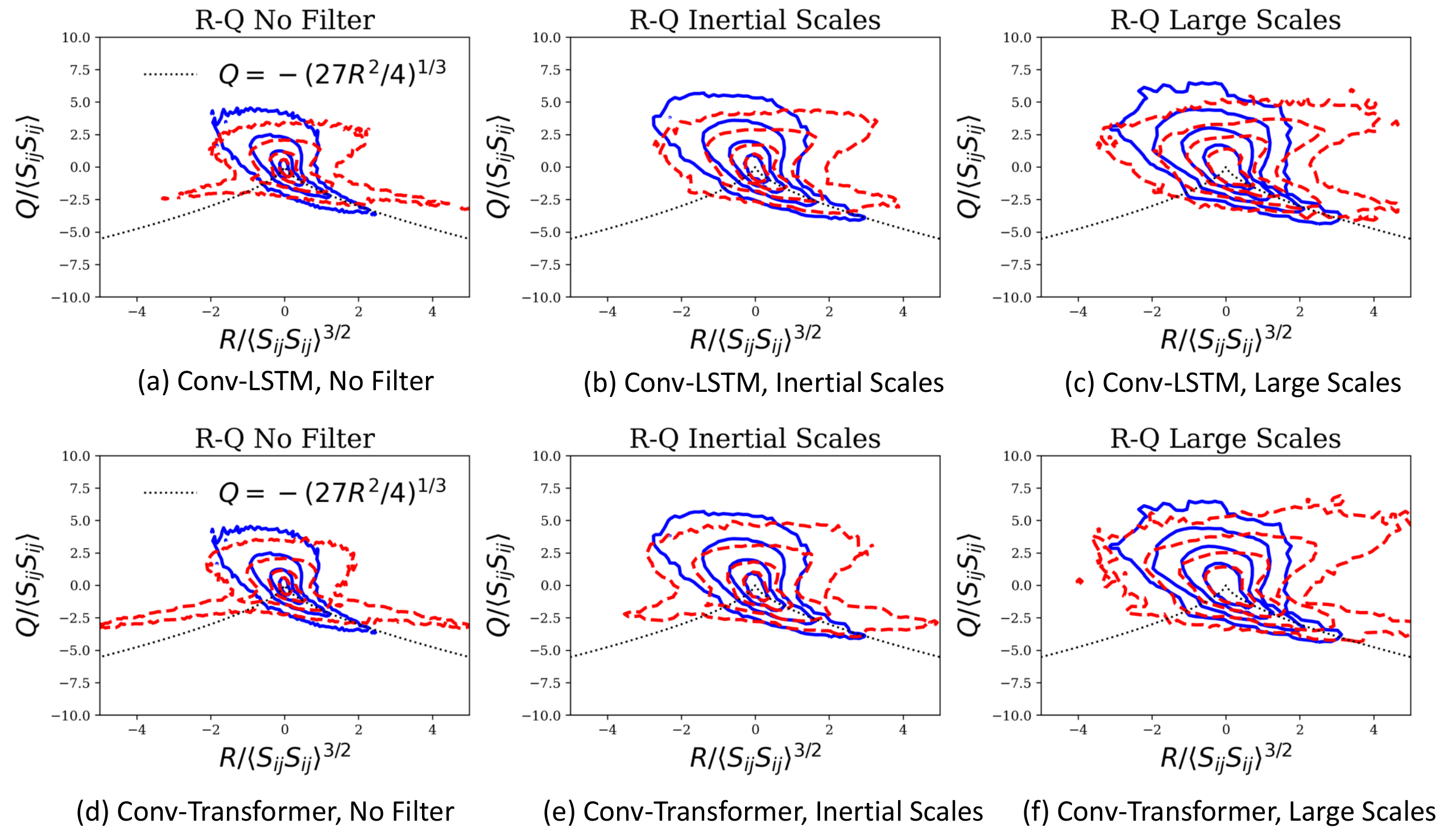}
 \caption{Predictions of R-Q using (a-c) Conv-LSTM and (d-f) Conv-Transformer models on small (a,d), inertial (b,e), and large (d,f) scales.}
 \label{fig:RQ}
\end{figure*}

During short-term predictions the models receive input sequence of $[X_{t-2b}, X_{t-b}, X_{t}]$ and generate sequence of $[\widehat{X_{t+b}}, \widehat{X_{t+2b}}, \widehat{X_{t+3b}}]$, where $\widehat{X}$ represents prediction of $X$. It is worth pointing out that we 
obtained robust results over all test data and the following results are for the test case where $t=6.8 T_{L}$.
In all of the diagnostics tests, we observe that the accuracy of results deteriorates from the first to the third predicted snapshots. This is quite expected as the error propagates from the first predictions to the next ones due to autoregressive nature of the predictions. 
Indeed, during the inference we obtain $\widehat{X_{t+b}}$, $\widehat{X_{t+2b}}$, and $\widehat{X_{t+3b}}$ based on $[X_{t-2b}, X_{t-b}, X_{t}]$, $[X_{t-b}, X_{t}, \widehat{X_{t+b}}]$, and $[X_{t}, \widehat{X_{t+b}}, \widehat{X_{t+2b}}]$, respectively.

In Figure~\ref{fig:VelocityComponents}, we evaluate the performance of our Conv-LSTM and Conv-Transformer models in reconstructing 2d snapshots (randomly sampled) of the velocity field, as well as the PDFs of the velocity components (where the statistics are based on the full 3d domain) across the predicted snapshots. The results show that for the first prediction time step both models capture reasonably well, Conv-LSTM is slightly better than Conv-Transformer, the instantaneous spatial structure of the flow and the statistical properties of the velocity field. Although the accuracy of reconstructed snapshots decreases for the next predictions due to the error propagation, we clearly observe that the quality of predicted snapshots using Conv-Transformer is much better.

The turbulent kinetic energy (TKE) spectra of the predicted time instants are shown in Figure~\ref{fig:TKE} for the Conv-LSTM and Conv-Transformer models.
While the reconstruction quality decreases for the Conv-LSTM model from the first to the third predictions, our Conv-Transformer model accurately captures the large and inertial scales of flow, both quantitatively and qualitatively, with significant loss of information for the smallest scales.

The PDFs of the longitudinal and transverse components of the velocity gradient tensor for different filtering lengths are shown in Figure~\ref{fig:VelocityGradient} across different predictions of the Conv-LSTM and Conv-Transformer models. The results illustrate that our models across all the predicted time instants can accurately capture the body of these PDFs, but fail at retaining heavy tails and skewness. Our results show that the Conv-Transformer model has a better performance than Conv-LSTM one. Furthermore, we observe that the quality of the results improves as we move from inertial to large scales, indicating that loss of information mainly occurred at small scales.  

%% VelGrad Short-Term Predictions

The capability of our models in reconstructing the joint-PDF of the $Q$ and $R$ invariants of velocity gradient tensor across different predictions and flow scales are shown in Figure~\ref{fig:RQ} for the Conv-LSTM and Conv-Transformer models. Our models struggle to capture the behavior of the $Q,R$ PDFs and can only capture some of the most frequent characteristics of the flow (interior contours of the PDF). Such large discrepancies in recovering these joint PDFs may seem surprising given that our models were able to accurately capture the body of the PDFs of velocity gradient tensor. However, the invariants $Q,R$ depend not only upon the properties of the individual velocity gradient components, but also upon more subtle features such as the geometric alignments between the strain-rate and vorticity fields which our models fail to capture from the data.

In summary, we can conclude that the Conv-Transformer model outperforms the Conv-LSTM model and can better retain statistics of flow field across the entire prediction horizon. The outperformance of Transformer model can be heavily attributed to its ability to process input sequences as a whole rather than element by element which is typical in LSTM model. Such a characteristic enables Transformer to capture more faithfully the temporal correlations between sequence elements. 

we also tried to test our models, which have been trained for short-term prediction, on challenging tasks of varying sampling interval and long-term predictions. Not shown here, but our results indicate that the quality of reconstructed flow fields across all the predicted snapshots deceases for the varying sampling interval inference. Furthermore, we observed that during cyclic prediction (to generate long-term predictions), error (from one prediction to the next one) propagates significantly and accumulation of such errors repeatedly result in flow snapshots that do not retain important characteristics of turbulence.
\vspace{-0.2cm}
\section{Conclusions}\label{Conclusions}
\vspace{-0.1cm}
In this study, we aim at exploring the feasibility of emulating temporal and spatial patterns of three-dimensional isotropic turbulence, purely from data, via modern deep learning approaches. Since our data size is memory intensive, we first generate a low-dimensional representation of the velocity data and then pass it to a sequence prediction network that learns the spatio-temporal correlations of the underlying data. Our results show that the accuracy of results for both models deteriorates across predicted snapshots due to autoregressive nature of the predictions. One potential idea to improve the paper is by including explicit multi-resolution structure in the latent code generation~\cite{jia2019recurrent}.

Based on our diagnostics tests, the Conv-Transformer model outperforms the Conv-LSTM one and can accurately, both quantitatively and qualitatively, retain the large scales, capture well the inertial scales of flow but fail at recovering the small and intermittent fluid motions. The outperformance of Conv-Transformer model can be heavily attributed to its ability to process input sequences as a whole rather than element by element (which is typical in LSTM model). However, both models neglect important characteristics of the joint-PDF of the Q and R invariants of velocity gradient tensor indicating that our models struggle to capture more subtle features of velocity field such as the geometric alignments between the strain-rate and vorticity fields. 

\vspace{-0.2cm}
\section*{Acknowledgements}
This work was  supported by National Science Foundation (NSF) under Grant No. ACI-1548562 and the Office of Naval Research (ONR) under Grant No. N00014-18-1-2244.
\vspace{-0.4cm}

\bibliography{References}

% \appendix
% \section{Acknowledgments}
% This work used the Extreme Science and Engineering Discovery Environment (XSEDE), which is supported by National Science Foundation grant number ACI-1548562 \cite{xsede}. Specifically, the Comet cluster was used under allocation CTS170009 and the authors would like to thank Marty Kandes for his assistance with setting up the GPU environment. This work was also supported by the Office of Naval Research (ONR) under grant number N00014-18-1-2244.

\end{document}